\documentclass{aa}
\usepackage{graphicx}
\usepackage{txfonts}
\usepackage{natbib}
\bibpunct{(}{)}{;}{a}{}{,} % to follow the A&A style

\begin{document}

\title{A three-dimensional picture of the delayed-detonation model of 
Type Ia supernovae} 

\author{E. Bravo\inst{1,2} \and  D. Garc\'\i a-Senz\inst{1,2}} 

\offprints{E. Bravo}

\institute{Departament de F\'\i sica i Enginyeria Nuclear, UPC, 
Jordi Girona 3,  M\`odul B5, 08034 Barcelona, Spain 
\and 
 Institut D'Estudis Espacials de Catalunya, Gran Capit\`a 2-4 08034 Barcelona\\
\email{eduardo.bravo@upc.edu\qquad domingo.garcia@upc.edu}
} 

\date{Received .....2007 / Accepted ....?}

\abstract
{}
{Deflagration models poorly explain the observed diversity of SNIa. 
Current multidimensional simulations of SNIa predict a 
significant amount of, so far unobserved, carbon and oxygen moving at low
velocities. It has been proposed that these drawbacks can be resolved
if there is a sudden jump to a detonation (delayed detonation), but these kinds
of models have been explored mainly in one dimension. 
Here we present new three-dimensional
delayed detonation models in which the
deflagraton-to-detonation transition (DDT) takes place in conditions like those
favored by one-dimensional models.}
{We have used a smoothed-particle-hydrodynamics code 
adapted to follow all the dynamical phases of the explosion, with algorithms
devised to handle subsonic as well as supersonic combustion fronts. The 
starting point was a centrally ignited C-O white dwarf of 
$1.38\mathrm{M}_{\sun}$. When the 
average density on the flame surface reached $\sim2-3\times 10^7$~
g cm$^{-3}$ a detonation was launched.}
{The detonation wave processed more than 0.3~M$_{\sun}$ of carbon and oxygen,
emptying the central regions of the ejecta of unburned fuel and raising its
kinetic energy close to the fiducial $10^{51}$~ergs expected from a healthy Type
Ia supernova. The final amount of \element[][56]{Ni} synthesized also was in the correct
range. However, the mass of carbon and oxygen ejected is still too high.}
{The three-dimensional delayed detonation models explored here 
show an improvement over pure deflagration models, but they still fail to
coincide with basic observational constraints. However, there are many 
aspects of the model that are still poorly known (geometry of flame ignition, 
mechanism of DDT, properties of detonation waves traversing a mixture of fuel
and ashes). Therefore, it will be worth pursuing its exploration to see if a 
good SNIa model based on the three-dimensional delayed detonation scenario can 
be obtained.}  


\authorrunning{Bravo \& Garc\'\i a-Senz}
\titlerunning{A 3D-picture of the delayed detonation model of SNIa}

\maketitle

\section{Introduction}

The knowledge of the physical mechanism by which a white dwarf is disrupted 
by a thermonuclear explosion is relevant to many topics of modern 
astrophysics. 
A satisfactory model of the explosion becomes crucial to better 
understand Type Ia supernovae, which in turn 
have profound implications in cosmology and in studies of the dynamics 
of the interstellar medium and the chemical evolution of galaxies. 
Nevertheless the details of the explosion mechanism are not well known.
 Nowadays 
the best models involve a white dwarf with a mass near the Chandrasekhar-mass
 limit 
in which a nuclear fuel ignites in one or many sparks around the center and
thermonuclear combustion propagates rapidly 
through the rest of the star. In these models the starting point of  
hydrodynamical 
calculations is the thermal runaway 
caused by the screened \element[][12]{C}+\element[][12]{C} nuclear reaction in degenerate 
conditions. 
The shape of the initially incinerated region probably does not keep the 
spherical 
symmetry that characterizes 
previous, hydrostatic, stages due to the stochastic nature of ignition and to 
the role played by 
hydrodynamic instabilities. As a consequence, any comprehensive numerical 
simulation of the explosion 
has to be performed in three dimensions. Many exploratory 
studies carried out in the past using one-dimensional hydrodynamics have shown
 that the  
key point to achieve a successful explosion relies in the proper estimation  of 
the velocity of the combustion front \citep{nie97,hil00}. 
It has been shown \citep{tim92} that the width of the flame is about $1 \mu$m at
$\rho\simeq 10^9$~g cm$^{-3}$, implying that at scales of a km the 
flame can be regarded as a thermal discontinuity separating ashes from fresh 
nuclear fuel. Although   
the flame velocity is large, close to 100~km~s$^{-1}$, 
it is quite subsonic:
$\simeq 0.01 v_\mathrm{sound}$, where $v_\mathrm{sound}$ is the sound speed. These velocities are   
far below the nearly supersonic velocities which, according to many 
one-dimensional simulations, 
are needed to blow away the star. It is today widely accepted that the
hydrodynamic instabilities inherent to the explosion 
are the drivers of combustion acceleration, thanks to the huge increase in the 
effective area of heat exchange between fuel and ashes.

Up to now the majority of three-dimensional studies of Type Ia supernovae have
 focused on the  pure deflagration paradigm
 \citep{gar98,rei02,gam03,gar05,sch06a,sch06b,rop06a,rop06b}.
 In these models the flame front
 moves fast but always subsonically, leaving time for the white dwarf to expand 
 before the deflagration reaches the outermost regions of the star. These  
 models have been successful in explaining many observational facts 
using a minimum set of free parameters. Unfortunately, all the 
three-dimensional versions of the pure deflagration paradigm present several
weaknesses:
\begin{itemize}
\item The presence of a few tenths of unburnt, and unseen, carbon moving 
at low velocity is probably the worst flaw \citep{koz05}.
\item Although the amount of Fe-group elements synthesized in the explosion is
within the expected range, the actual mass of \element[][56]{Ni} ejected might 
turn out to be too
low when neutronization due to electron captures is properly taken into account.
\item The final kinetic energy never reaches the canonical value of 1~foe 
(1 foe $\equiv 10^{51}$~ergs).
\item The synthesis of intermediate-mass elements is scarce 
\citep[see][]{gar07}.
\item The ejecta are not chemically stratified, as demanded by observations of supernovae and remnants \citep{bad06,ger07}.
\item Large clumps of radioactive \element[][56]{Ni} and \element[][56]{Co} are present at the photosphere 
at the time of maximum brightness.
\end{itemize}

There have been several attempts to bypass the weaknesses of deflagrations
 models, 
preserving at the same time many of their advantages, by postulating a sudden  
transition from a subsonic deflagration to a supersonic detonation
\citep{iva74}, i.e. 
the multidimensional version of the delayed detonation model:
\citet{kho91} in 1D; \citet{arn94,liv99,gol05} in 2D;
\citet{gar03,gam04,gam05,rop07} in 3D. In particular, the recent work of 
 \citet{gam05} 
gave a satisfactory thermonuclear explosion solving
 many of the problems of pure deflagration models.  
The density at which the jump to the detonation takes 
place had to be 
estimated because the physics behind the deflagration-to-detonation (DDT)
transition is not well understood.
In \citet{gam05} the adopted transition density was
rather high, $\rho_\mathrm{t}\ge 2.5\times 10^8$~g cm$^{-3}$ and the detonation 
was induced near the center of the white dwarf. 
\citet{rop07} explored the range of explosions that 
would result if the
DDT took place in the outer shells, at $r\sim1.7-1.8\times10^8$~cm, after a huge 
expansion drived by subsonic combustion initiated in a few hot 
bubbles. They used a physically motivated criterion
to switch on a DDT, namely that the combustion entered the distributed regime, 
in which mixing of burnt and fresh fuel by turbulence fluctuations at the scale 
of the flame width might trigger a detonation. Although this criterion is as fragile as any other
currently used in multidimensional simulations of DDT, it led to consistent results
with respect to the final kinetic energy and to the mass of Fe-group and
intermediate-mass elements synthesized. It also allowed almost complete 
depletion of carbon in the center. 

Given the scarcity of three-dimensional simulations of the delayed
detonation scenario and the complexity of the problem, it is interesting to
perform additional calculations to explore further the range of initial 
conditions of the DDT. 
In this work we present new three-dimensional simulations 
of delayed detonations in white dwarfs in which the DDT
takes place in conditions close to those usually assumed in one-dimensional
calculations. In the present models the DDT was triggered when the mean density at the flame
surface was $\rho\sim2-3\times 10^7$~g cm$^{-3}$. We have also explored the
sensitivity of the results to the initial geometrical configuration of the flame
by computing a delayed detonation model in which the deflagrative phase started
at the center of the white dwarf, and another model in which it started 
in 30 hot bubbles, as in the deflagration model B30U in \citet{gar05}.

The plan of the paper is as follows.
In Sect. 2 we briefly describe the hydrocode and the physics included in the 
three-dimensional simulations. Next, we discuss the 
hydrodynamical evolution during both the deflagration and the detonation stages,
as well as the final nucleosynthesis. Some discussion of our
results and a comparison to previous works as well as the conclusions are 
provided in Sect. 4. 

\section{Methods}

The simulations were carried out using a smoothed-particle hydrodynamics (SPH) 
code suited to
 handle both subsonic deflagration waves \citep{gar98} and 
 supersonic detonations \citep{gar99}. The
 calculations were carried out using 250\,000 particles, which translated into a
 maximum spatial resolution of 22~km\footnote{Here we give the smoothing length
 of the SPH code as a proxy for the spatial resolution. Note that the smoothing
 length is both spatially and temporally variable} (about 1\% of the white dwarf
 radius) during the deflagration phase and, later on, of 15~km ($\sim0.2\%$ of
 the stellar radius) during the detonation phase. 
 The equation of state, the same as described in \citet{gar05}, includes all 
 the terms relevant for the explosive phases of Type Ia supernovae. The 
 properties of matter in nuclear statistical equilibrium (NSE) were
 interpolated from accurate tables giving the nuclear binding energy, electron
 capture and neutrino generation rates as a function of the temperature, density
 and electron mole number. This represented a negligible overhead of CPU time
 and allowed us to obtain realistic estimates of the nucleosynthesis during the
 SNIa explosion. As usual in this kind of simulation, the 
detailed nucleosynthesis was calculated by postprocessing the output of the 
hydrodynamical model. 
 
 During the
 deflagration phase the flame was propagated by a diffusion-reaction scheme
 devised to match a prescribed flame velocity law. Although the SPH algorithms are flexible enough to incorporate a subgrid-scale model able to capture the effects of turbulent flows into the effective flame velocity (Garc\'\i a-Senz et al.\ 2008, in preparation), for the present calculations
 we choose to stay as close to possible to the usual assumptions of succesful
 one-dimensional delayed detonation models. Thus, we implemented a
 constant flame velocity in the range $v_\mathrm{flame}=100-200$~km s$^{-1}$ , independent of density, during the deflagration phase. As described in
 \citet{gar05}, this prescription has proven adequate to describe the gross 
 features of subsonic burning propagation. Note that the true laminar velocity 
 in the center of the white dwarf is close to 
 $v_\mathrm{flame}\simeq 100$~km s$^{-1}$ . On the other hand the time-averaged  
effective deflagration velocity which stems from that scheme is similar to the
one assumed in successful one-dimensional delayed detonation 
models during the 
deflagration phase , $\simeq 0.03v_\mathrm{sound}-0.05v_\mathrm{sound}$ 
\citep{dom00}.

During the detonation phase the above described algorithm for flame propagation
was switched off and the combustion was followed by solving the nuclear kinetic
equations together with the hydrodynamic and energy equations. The nuclear part
consisted of a small nuclear network of 9 nuclei from helium to nickel
\citep{tim00} with prescriptions aimed to handle both quasi-statistical
equilibrium and nuclear statistical equilibrium. Whenever the temperature 
became higher than 
$5.5\times10^{9}$~K the material was assumed to be incinerated to NSE if the
density was higher than $10^6$~g cm$^{-3}$. Once the NSE regime was attained the
abundances were not computed by solving the nuclear kinetic equations, instead the chemical composition was obtained from the NSE tables. In our calculations, incinerated particles were assumed to leave NSE when either their temperature or their density were lower than
$2\times10^9$~K or $10^6$~g cm$^{-3}$, respectively. During this phase
the detonation front was captured, that is the particles within the detonation 
front were identified with the aid of an algorithm that looked for 
simultaneous strong gradients of pressure and temperature together with sonic flows. Particles located at a detonation front were assigned  
ellipsoidal kernels to perform the SPH interpolations in such a way that
the smallest axis of its ellipsoid was aligned with the normal to the
detonation front. In regions where there were no shocks the kernel reduced to
the standard spherically symmetric kernel. The ability of this device to
represent detonation waves was demonstrated in Fig. 1 of \citet{gar99}, where 
more details of the numerical method are available. Typically,
these techniques lead to an improvement of a factor $\sim2-3$ in spatial resolution. 

The initial model consisted of an isothermal C-O white dwarf of $1.38~\mathrm{M}_{\sun}$ in hydrostatic equilibrium, 
with a central density $\rho_\mathrm{c}=1.8\times 10^9$~g cm$^{-3}$. We ran two
simulations of the delayed detonation scenario, whose main features are given in
Table \ref{tab1}. 
In model DDT3DA the first stages of the explosion were followed using a 
one-dimensional Lagrangian code until the mass burnt was 
$0.09~\mathrm{M}_{\sun}$ and the central density had declined to 
$\rho_c=1.4\times 10^9$~g cm$^{-3}$. 
At that point the structure was mapped to a 3D distribution of 250\,000 
particles, and the radial velocity field at the flame location was perturbed by a sinusoid of wavenumber $l=20$ and amplitude $200$~km s$^{-1}$, as
much as the velocity of matter behind the flame at that time. The ensuing 
evolution was followed with the three-dimensional SPH code, with a baseline velocity of $v_\mathrm{flame}=100$~km~s$^{-1}$.
The initial geometry of the flame in model DDT3DB was different,
consisting in a set of 30 sparks scattered around the center of the 
white dwarf. The three-dimensional calculation of this model started 
exactly as in model B30U of \citet{gar05}, to which 
we refer the reader for more details.

The switch from the subsonic combustion front to the supersonic regime also was handled in a different way in both models. In DDT3DA the detonation was induced
simultaneously in many points on the flame surface, when its average density declined 
to $\sim2\times 10^7$~g cm$^{-3}$. By contrast, in model DDT3DB the detonation was 
artificially triggered in a single region of the flame, although at a density 
similar to that in DDT3DA.

\begin{figure}
\resizebox{\hsize}{!}{\includegraphics{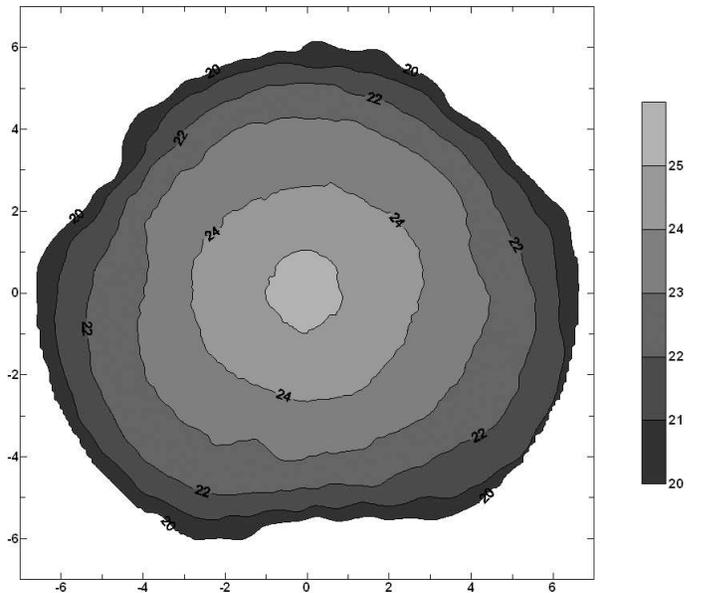}}
\caption{
Contour map of pressure in a cut across the star at the end of the deflagration 
phase of model DDT3DA ($t=1.55$~s). The labels of the isobars give the logarithm
of pressure in erg cm$^{-3}$. The axis labels are in units of 1\,000 km. The 
preservation of spherical symmetry is a consequence of the subsonic nature of 
the deflagration front.
}\label{fig4}
\end{figure}

The last three columns of Table~\ref{tab1} give details of the performance of
the code with respect to the conservation of total energy, momentum and angular
momentum. The conservation of the two last is excellent, whereas that of energy is
acceptable as the error stays around 1\% of the final kinetic energy. Also 
supporting the ability of the code to handle the subsonic deflagration phase 
is the spherical symmetry retained by the pressure field during the huge  
expansion of the white dwarf in this phase, as shown in 
Fig.~\ref{fig4}. Such a smooth 
pressure distribution has to be compared with the convoluted temperature field
depicted in the last snapshot of Fig.~\ref{fig1}. The limitations of our 
methodology to carry out simulations of Type Ia supernovae were widely 
discussed in \citet{gar05}.
 
\begin{table*}
\caption[]{Main features of the models.}
\label{tab1}
\centering
\begin{tabular}{llccccccc}
\hline\hline
\noalign{\smallskip}
Model & Ignition & $E^{\mathrm{a}}$ & $M \left(^{56}\mathrm{Ni}\right)$ &
$t_\mathrm{last}$ & $I$ & $\Delta E^{\mathrm{b}}$ &
$\Delta p^{\mathrm{c}}$ & $\Delta L^{\mathrm{d}}$ \\  
& & ($10^{51}$~erg) & ($M_\odot$) & 
(s) & (Time steps) & ($10^{51}$~erg) & ($10^9$ M$_{\sun}$ cm s$^{-1}$) & 
($10^{17}$ M$_{\sun}$ cm$^2$ s$^{-1}$) \\ 
\noalign{\smallskip}
\hline
\noalign{\smallskip}
DDT3DA & Central volume & 
0.75 & 0.54 & 7.9 & 7\,400 & $0.01$ & $4\times10^{-8}$ & $2\times10^{-8}$ \\
DDT3DB & 30 bubbles & 
0.81 & 0.63 & 463. & 21\,200 & $-0.008$ & $1\times10^{-8}$ & $10^{-8}$ \\
\noalign{\smallskip}
\hline
\end{tabular}
\begin{list}{}{}
\item[$^{\mathrm{a}}$] Final energy (gravitational+internal+kinetic) of the
ejected mass at the last computed time, $t_\mathrm{last}$.
\item[$^{\mathrm{b}}$] Accumulated error in the total energy at the last 
computed model, after $I$ time steps.
\item[$^{\mathrm{c}}$] Total momentum at the last computed model, normalized
taking a characteristic velocity of the ejecta of $10^9$~cm~s$^{-1}$.
\item[$^{\mathrm{d}}$] Total angular momentum at the last computed model,
normalized taking a characteristic velocity of the ejecta of 
$10^9$~cm~s$^{-1}$, and a characteristic radius of $10^8$~cm.
\end{list}
\end{table*}

\section{Results of the three-dimensional simulations} 

\subsection{Model DDT3DA: The deflagration phase}

This phase begins when the first sparks ignite 
in the central regions of the white dwarf and ends when the 
first detonation wave appears. Both modes of burning propagation can
coexist for a while during the detonation phase although the efficiency of the
 deflagration is far below that of the supersonic detonation. As mentioned
 before, in our computations the deflagration algorithm was switched off once a
 DDT was artificially induced. \citet{rop07} have discussed the impact of subsonic 
 burning after a DDT. They found that it is relevant only when the mass burnt 
 during the deflagration phase is quite low as is the case, for instance, if
 the runaway starts in a few hot bubbles.

The geometry of the initial sparks which light the core of the white dwarf 
is not known. However, in the three-dimensional realm it is to be expected that
the ignition starts asynchronously in a number of small regions 
scattered around the core \citep{gar95,wun04,kuh06}. For a large enough number 
of igniting points located close to the center, bubbles of incinerated 
matter would merge before they 
have had the opportunity to detach from the burning pack. Were this the case, 
the hydrodynamical evolution 
would not be very different from that due to the massive ignition of the central
volume of the core. This is the scenario we examine in model DDT3DA, in which 
the explosion was triggered by incinerating a small amount of mass in the 
central part of the core of an isothermal Chandrasekhar-mass white dwarf. 

\begin{figure*}
\centering
\includegraphics[width=17cm]{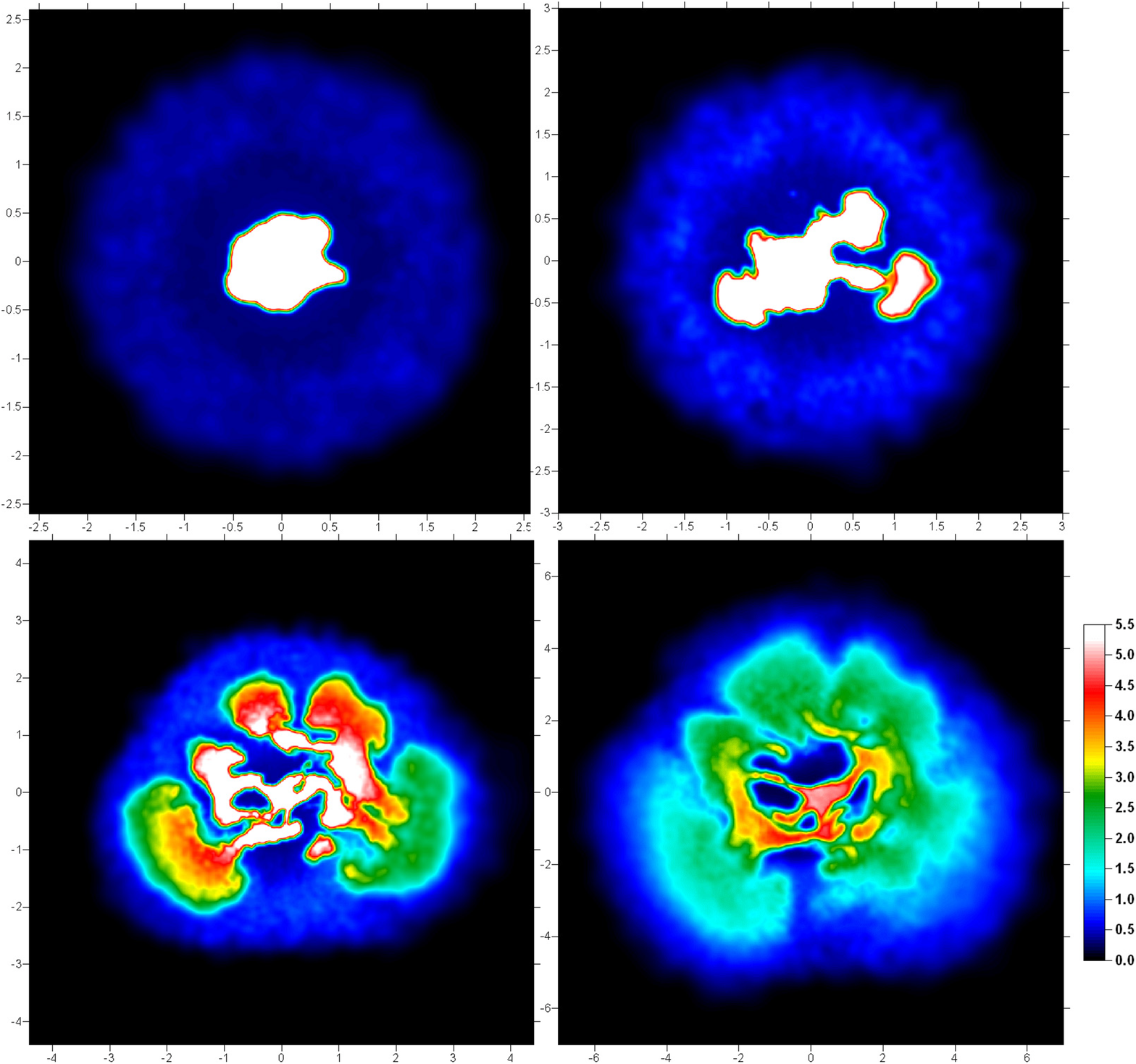}
\caption{
Snapshots of the temperature map in a cut across model DDT3DA at
times $t=0.56$, 0.86, 1.20, and 1.55~s, all of them belonging to the
deflagration phase. The lengthscale is given in units of $1\,000$ km s$^{-1}$,
while the temperature scale is given in the side colorbar in units of $10^9$ K.
}\label{fig1}
\end{figure*}

\begin{figure}
\resizebox{\hsize}{!}{\includegraphics{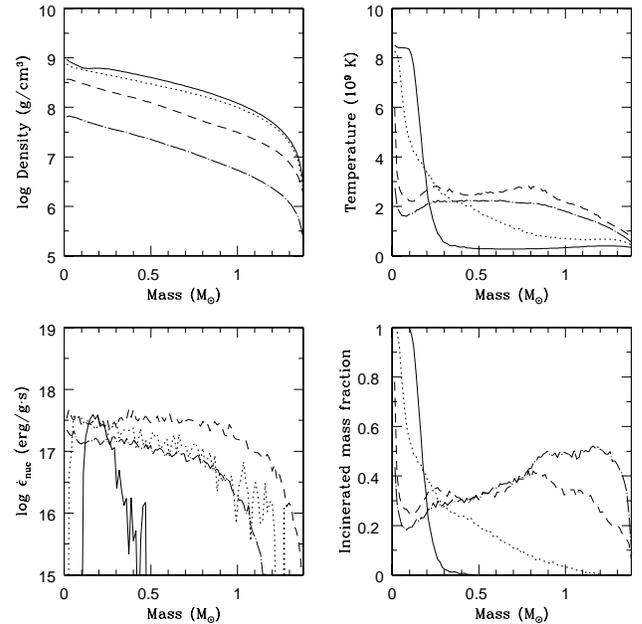}}
\caption{
Angle-averaged profiles of density, temperature, nuclear energy generation, and
incinerated mass fraction during the deflagration phase. The lines
represent the profiles at the same times as Fig.~\ref{fig1}: $t=0.56$~s (solid),
0.86~s(dotted), 1.20~s (short dashed), and 1.55~s (dot-long dashed).
}\label{fig2}
\end{figure}

\begin{figure}
\resizebox{\hsize}{!}{\includegraphics{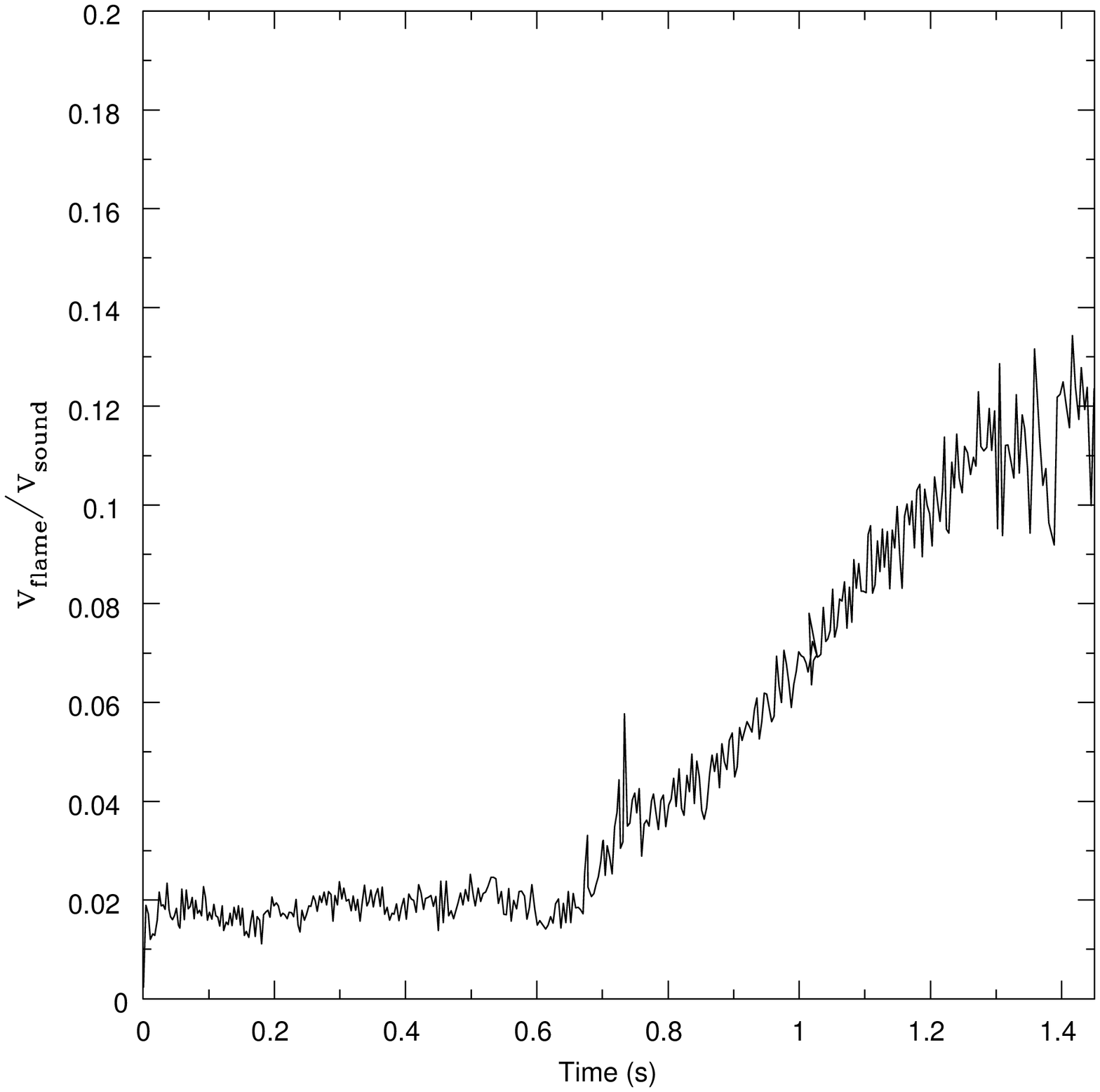}}
\caption{
Effective velocity of the deflagration front in units of the local sound 
speed as a function of time.
}\label{fig3}
\end{figure}

The evolution of the star during the deflagration phase can be seen in 
Figs.~\ref{fig1} and \ref{fig2}. In Fig.~\ref{fig1} there are several 
snapshots of the evolution from the beginning of the SPH calculation, which we 
define as $t=0$~s, when the flame surface was completely spherical, up to 
$t=1.55$~s, when the flame was strongly distorted by the Rayleigh-Taylor 
instability and the average density at the flame was 
$<\rho_\mathrm{flame}>\simeq2\times 10^7$~g cm$^{-3}$. A complementary view 
is given in Fig.~\ref{fig2} where we plot the angle-averaged radial profiles of 
density, temperature, nuclear energy generation rate, and incinerated mass 
fraction during the deflagration phase. During the first 
half a second the flame front remained nearly spherical in spite of the
sample of perturbations of the velocity field that were seeded at $t=0$~s. 
Afterwards, the fingers characteristic of the Rayleigh-Taylor instability 
developed very quickly (second and third snapshots in Fig.~\ref{fig1}), which
resulted in hot 
material rising and cool unburnt carbon and oxygen intruding towards the 
center (fourth panel of Fig.~\ref{fig2}). As the flame surface increased due to the hydrodynamic instabilities, the rate of released nuclear energy also did. The effective burning velocity 
kept pace with the evolution of the surface, as can be seen in Fig.~\ref{fig3}:
it remained around $0.02 c_\mathrm{s}$~until $t=0.6$~s and rose monotonically up
to $0.14 c_\mathrm{s}$~at $t=1.4$~s. This large increase in the effective
burning rate when t$\geq$~0.6 s is as well apparent in the third panel of
Fig.~\ref{fig2} which depicts the nuclear energy generation rate. At $t=0.86$~s
(second snapshot in Fig.~\ref{fig1}) the burning front clearly displays the mushroom-like shape characteristic of the non-linear Rayleigh-Taylor regime, encompassing a large fraction of the white dwarf mass. As a consequence 
the rate of nuclear energy generation went up steeply until a maximum value 
of $1.3\times 10^{51}$~erg s$^{-1}$ was achieved at t=1.1~s. During the last 
second of deflagration propagation the average 
temperature and nuclear energy generation rate were uniform throughout the white dwarf. 

When the elapsed time was $\sim1.55$~s the average density of the flame became 
low enough to allow intermediate-mass elements to be synthesized. At this 
time the central density had declined to $\rho_\mathrm{c}\simeq 
5.4\times10^7$~g cm$^{-3}$ and the expansion was fast enough that the nuclear
reactions began freezing at the outermost regions of the deflagration front. The main features of the model at the end of 
this phase are summarized in Table~\ref{tab2}. The amount of nickel synthesized,
0.27~M$_{\sun}$, was too low to give a suitable explosion with the maximum
luminosity of a typical SNIa. In addition, there remained too much carbon and 
oxygen. 

\begin{table*}
\caption[]{Main features of the models at the time of DDT.}
\label{tab2}
\begin{tabular}{llcccccccc}
\hline\hline
\noalign{\smallskip}
Model & DDT & 
$t^\mathrm{DDT}$ & $\rho_\mathrm{c}^\mathrm{DDT}$ & 
$\rho_\mathrm{fl}^\mathrm{DDT}$ & $E_\mathrm{tot}$ &
$M_\mathrm{NSE}^\mathrm{DDT}$ & 
$M(^{56}\mathrm{Ni})^\mathrm{DDT}$ & $M(\mathrm{C-O})^\mathrm{DDT}$ \\ 
& & 
(s) & ($10^7$~g cm$^{-3}$) & ($10^7$~g cm$^{-3}$) & ($10^{51}$~erg) & 
(M$_{\sun}$) & (M$_{\sun}$) & (M$_{\sun}$) \\ 
\noalign{\smallskip}
\hline
\noalign{\smallskip}
DDT3DA & Multipoint &
1.55 &  5.4 & 2.0 & 0.40 & 0.50 & 0.27 & 0.65 \\
DDT3DB & Single point &
0.81 & 22.2 & 3.7 & 0.39 & 0.58 & 0.43 & 0.64 \\
\noalign{\smallskip}
\hline
\end{tabular}
\end{table*}
 
\begin{figure}
\resizebox{\hsize}{!}{\includegraphics{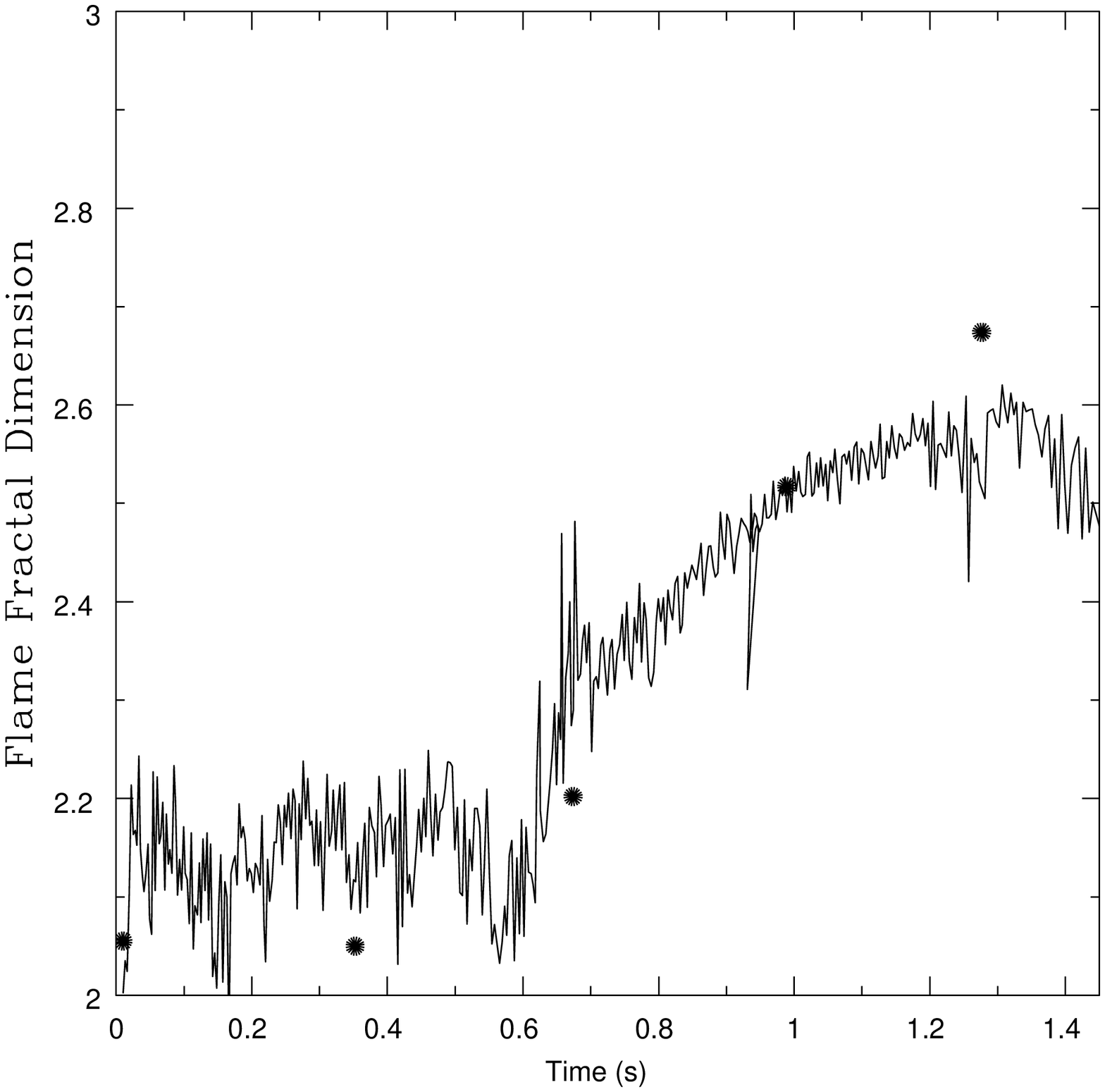}}
\caption{
Evolution of the fractal dimension of the flame. Dots show the correlation
dimension, while the solid line represents an alternative estimate of the 
fractal dimension derived from the effective rate of mass burning.
}\label{fig5}
\end{figure}

The scaling properties of the flame surface during the deflagration phase were
monitored, using the method described in \citet{gar98}, and the results are presented 
in Fig.~\ref{fig5} (filled dots). This kind of calculation, which gives the 
so-called correlation dimension, provides a purely geometric description of
the flame structure determined by the spatial distribution of particles 
belonging to the flame. An independent way of obtaining the fractal dimension 
is to estimate the total flame surface within the minimum 
lengthscale resolved by the code, $h$, and the integral scale, using the ratio of the mass consumption rate, $\dot M$, and the 
baseline flame velocity (solid line in Fig.~\ref{fig5}):

\begin{equation}
\centering
D=2+\frac{\log\left(\dot M/4\pi\bar r^2\rho~v_\mathrm{flame}\right)}
{\log\left(\bar{r}/h\right)}\,
\end{equation}

\noindent where the integral scale has been approximated by the average radius 
of the flame, $\bar{r}$. As it can be seen in Fig.~\ref{fig5} there is excellent agreement between both methods of computation of the fractal dimension, implying that the above equation can be safely used in analytical and one-dimensional calculations to explore the physics of thermonuclear supernovae. After an 
initial period where the fractal dimension remained close to 2.0 it 
rapidly began to increase after $t=0.6$~s, at which time the average density 
of the flame had dropped below $5\times 10^8$~g cm$^{-3}$~. Approximately 1.1~s after the beginning of the three-dimensional calculation the 
fractal dimension had reached a value $D\simeq 2.6$, similar to that 
associated with the RT instability in the non linear regime, 
$D_\mathrm{RT}=2.5$. That high value of the fractal dimension indicates that 
the Rayleigh-Taylor instability rather than turbulence\footnote{In the
discussion we are assuming Kolmogorov turbulence \citep{zin05,sch06b}, 
characterized 
by a fractal dimension $D_\mathrm{K}=2.33$, although the turbulence might also 
be of the Bolgiano-Obukov type, characterized by $D_\mathrm{BO}=2.6$ 
\citep{nie00}} was the dominant 
mechanism wrinkling the flame at those scalelengths directly resolved by the 
hydrocode. This was expected as our resolution is insufficient to
resolve turbulence, i.e. the minimum smoothing length attained in the SPH
calculation during deflagration was 20~km while the integral length of the
turbulent cascade is $\sim10$~km \citep{lis00}. Afterwards, the fractal 
dimension saturated at $D\simeq 2.6-2.7$.   

\begin{figure}
\resizebox{\hsize}{!}{\includegraphics{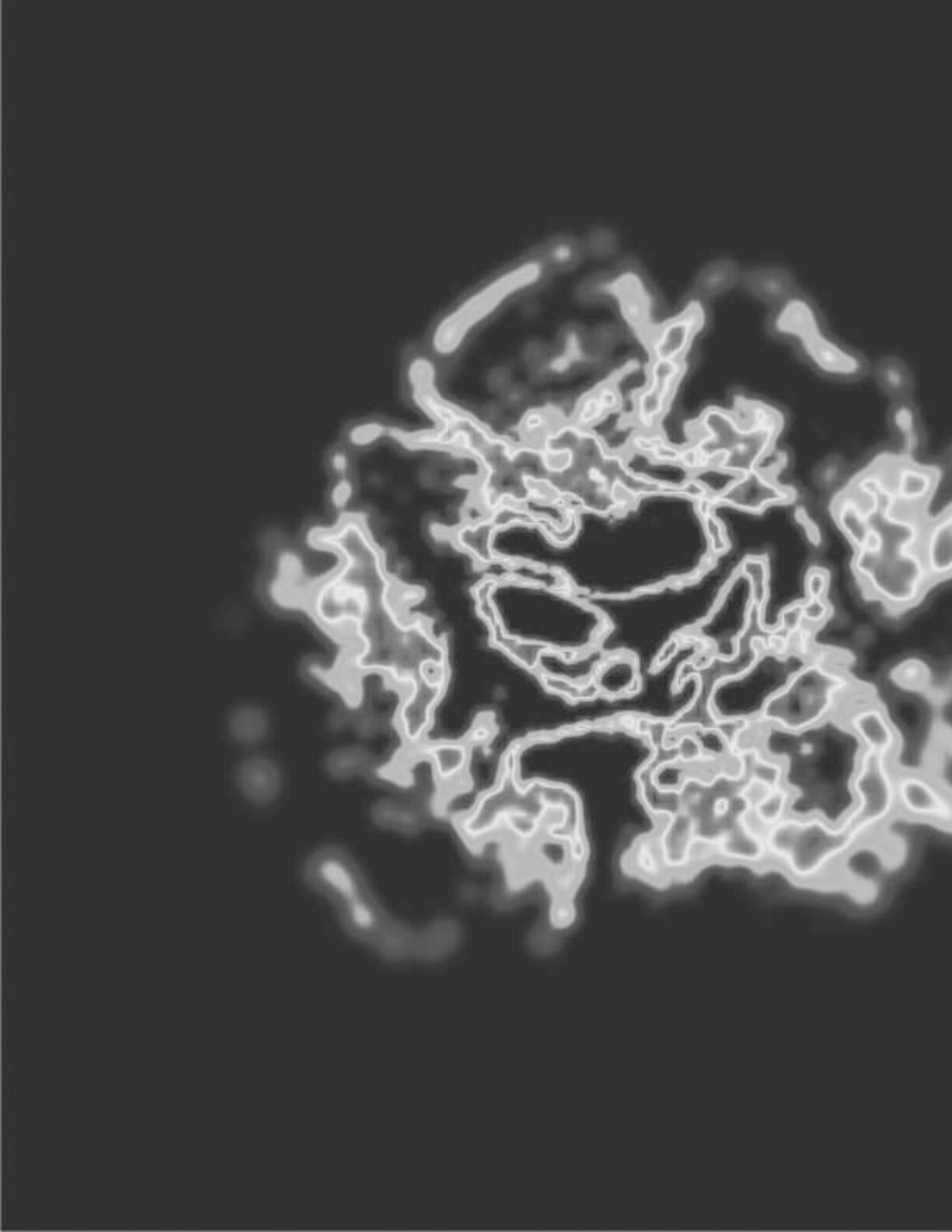}}
\caption{
Flame structure in a cut of the white dwarf at the end of the deflagration 
phase. The image shows the location of hot fuel whose temperature lies in the
range $1-3\times10^9$~K.
}\label{fig6}
\end{figure}

\subsection{Model DDT3DA: The detonation phase}

As has been discussed in Section 2, explosions driven purely by a
deflagration usually run into trouble when compared to the observational
properties of normal SNIa events for various reasons. One of them is the
difficulty of reaching explosions with energies of roughly $10^{51}$~ergs. Another serious objection is that current three-dimensional calculations always predict a strong mixing of elements in the ejecta. 
A possible way to circumvent these drawbacks is to assume that at some time the subsonic 
deflagration turns into a supersonic detonation at one or several points of the 
flame surface. Although several mechanisms have been suggested to account for 
such a transition, a definitive answer to this question is still 
missing. Therefore, the temporal and spatial coordinates at which the DDT takes 
place should be regarded as free parameters of the model \citep{liv99}. 

\begin{figure*}
\centering
\includegraphics[width=17cm]{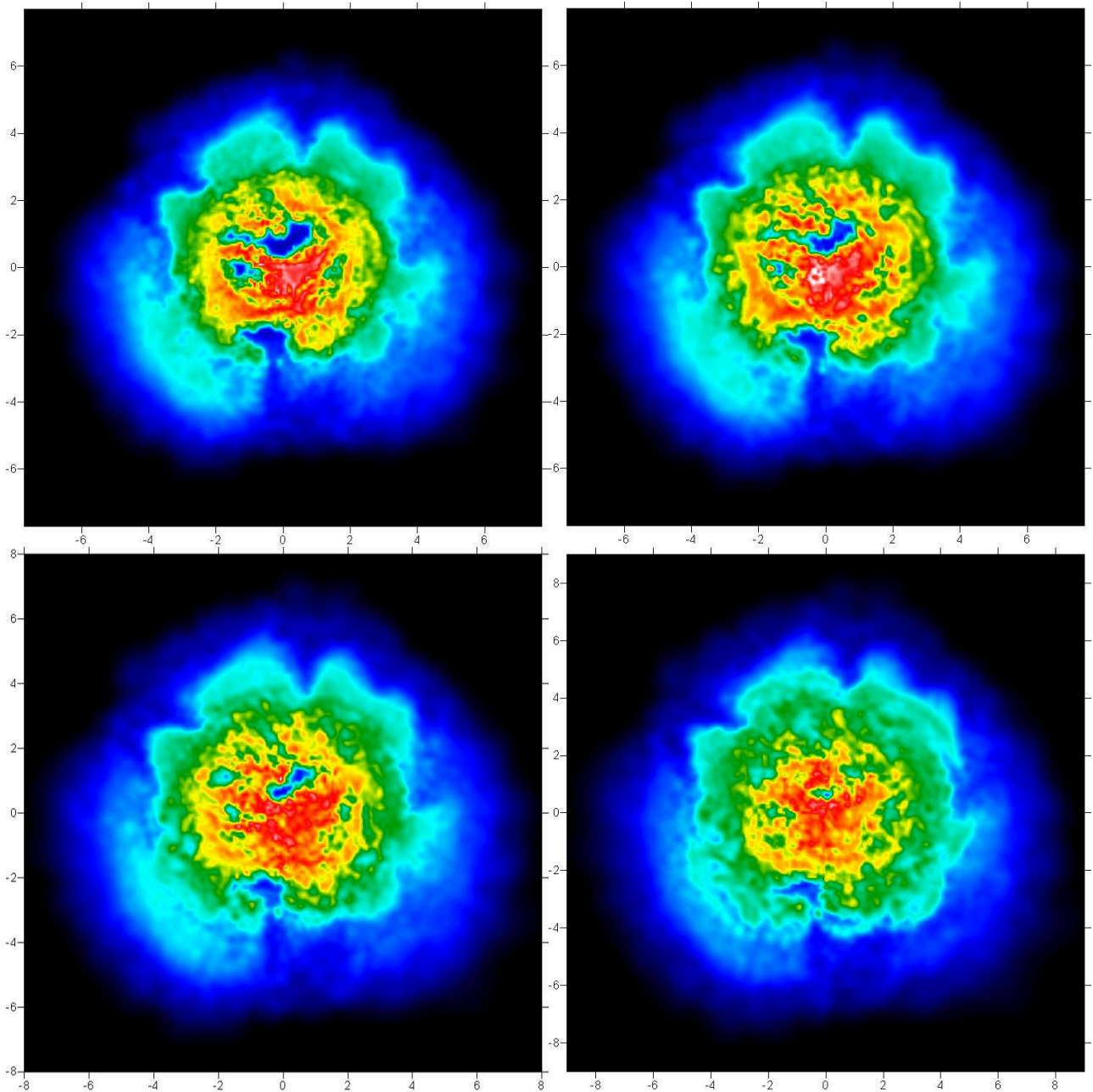}
\caption{
Snapshots of the temperature map in a cut across model DDT3DA 
at times $t=1.58$, 1.61, 1.66, and 1.75~s. The deflagration-to-detonation 
transition was induced just at 1.55~s. The scales of length and temperature are
the same as in Fig.~\ref{fig1}.
}\label{fig7}
\end{figure*}

Many one-dimensional simulations carried out in the nineties suggested that 
the agreement with observations is fairly good when the transition takes  
place at densities around $2-4\times10^7$~g cm$^{-3}$. This is about the 
average density of the flame when its fractal dimension stabilizes around 2.6, 
and approaches the density at which combustion is thought to leave the 
flamelet regime and enter into the distributed regime. Let us assume that the 
flame propagates with
the laminar velocity at the Gibson scale and that the flame surface behaves as 
a fractal of dimension 2.33 (a value determined by the turbulent energy spectrum) 
between the Gibson scale and the minimum scale resolved 
by the code, $h$. Then, the effective flame velocity can be obtained as a function of the fractal dimension, $D$, that characterizes the flame between $h$ and the integral scale. The result of this calculation is that the effective flame velocity exceeds the maximum 
velocity of a stable Chapman-Jouget deflagration \citep{kho88} if $D>2.5$ 
at $\rho\la2\times10^7$~g cm$^{-3}$. Thus, even though the laminar flame 
velocity always remains much lower than the sound velocity, the effective flame
velocity can be high 
enough to compress the material ahead, switching on a detonation \citep{woo94}.

As a practical criterion to select the particles prone to detonate
we flagged those placed at regions where the local fractal dimension was
higher than $D=2.5$ when the average density of the flame was
$<\rho_\mathrm{flame}>\simeq2\times 10^7$~g cm$^{-3}$. As these high-$D$ regions
trace the geometrical complexity of the flame it is not
unreasonable to assume that they can harbor conditions suitable for detonation initiation. 
In our calculation the above criterion was satisfied for the first 
time at $t=1.55$~s, marking the end of the deflagration phase. A two-dimensional
image of the flame structure at this time is shown in 
Fig.~\ref{fig6}, where  
its geometric complexity can be seen. The regions displaying the highest $D$ 
spanned the whole radial extent of the white dwarf, except for a narrow 
central volume and the outermost 0.2~M$_{\sun}$. Once the high-$D$ particles were selected, a detonation was artificially induced through their instantaneous incineration. Note that we do not mean that reaching a given fractal dimension of the flame 
is a necessary or sufficient condition for transition to detonation, but for 
our purposes it provided a convenient way to select the time and location of 
the DDT. 

\clearpage
\begin{figure}
\resizebox{\hsize}{!}{\includegraphics{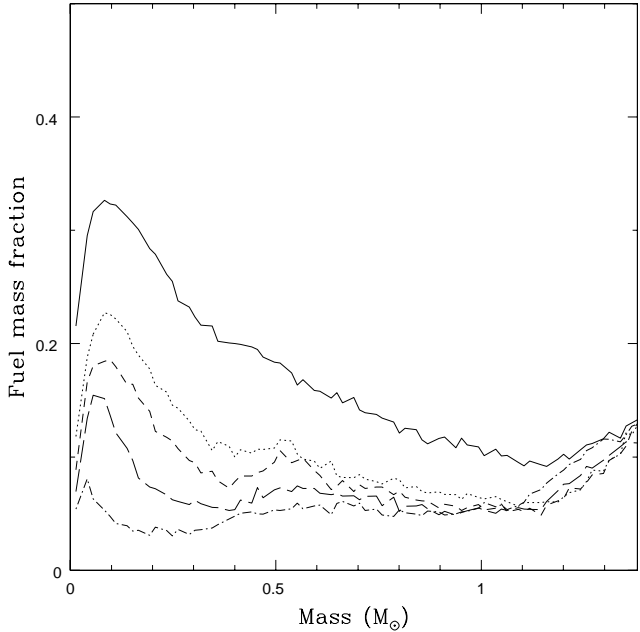}}
\caption{
Evolution of angle-averaged profiles of the \element[][12]{C}+\element[][16]{O} mass fraction 
during the detonation phase of model DDT3DA. The profile at DDT is shown together with those belonging to the same times shown in Fig.~\ref{fig7}.
}\label{fig8}
\end{figure}

\begin{figure}
\resizebox{\hsize}{!}{\includegraphics{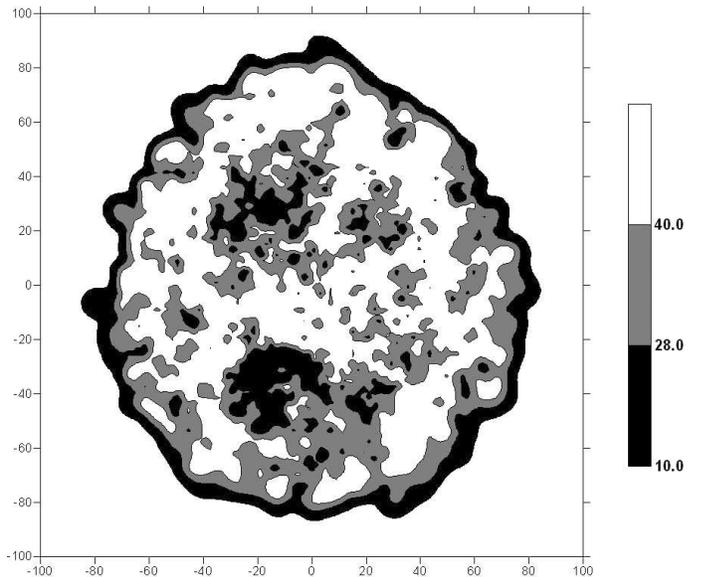}}
\caption{
Map of the final chemical composition on a cut across the ejected matter: 
C-O is shown in black,
intermediate-mass elements in grey, and Fe-Ni in white. The incinerated
matter pervades the whole ejecta, with intermediate-mass elements dominating an
annular region at intermediate radius. Unburned C-O are found in isolated
pockets, mainly in the outer regions of the ejecta.
The origin of the large C-O pockets seen in this figure can be traced back to features of the flame front identifiable in Figs.~\ref{fig1}, \ref{fig6}, and \ref{fig7}.
}\label{fig9}
\end{figure}

As we did with the deflagration phase, we describe here the evolution during the
detonation phase by means of two figures, Figs.~\ref{fig7} and \ref{fig8}, the 
first one showing several snapshots of the temperature field through a cut of 
the white dwarf, and the second one giving the evolution of the angle-averaged radial profiles of 
the fuel mass fraction. After a transient induction period, detonations settle in a 
steady state consisting of a shock plus a tiny reaction zone, 
followed by a much larger relaxation zone. Our code is only able to effectively 
resolve the last one. Thus, we cannot ascertain the result of the 
interaction of the detonation front with obstacles, i.e. inert volumes. 
However, matter incinerated to the NSE should not be described as 
inert, because at the densities we are considering it is able to react to the 
passage of a shock, that compress and heats matter, and readjust its 
composition accordingly 
\citep{mai06}. Hence, when the precursor shock of a detonation 
transits through NSE matter, it rises its temperature favouring lighter, less 
bound, species, which in turn depletes energy from the shock front. In our
calculations, the detonation did not survive the crossing of volumes composed of
nuclear ashes. The complex geometry of the flame at the time of DDT favored 
the formation of isolated pockets of fuel that could not be caught by 
the detonation front, so that they remained unburned (Fig.~\ref{fig9}). 

Because the detonation was initiated in a more or less uniform way throughout the white dwarf, there followed a period of rapid and generalized 
combustion leading to a temperature distribution much more homogeneus than that 
of the preceding deflagration phase (Fig.~\ref{fig7}). 
The inwards detonation wave synthesized an additional mass of
\element[][56]{Ni}, whereas the outgoing detonation, travelling through lower density
layers, synthesized intermediate-mass elements (Fig.~\ref{fig9}). 
In this way, the DDT allowed the partial compensation of the shortcomings of the 
previous deflagration stage, giving an improved explosion model. 

The main features of the explosion at the last computed time are given in 
Table~\ref{tab1}, while the nucleosynthesis obtained by post-processing the
thermal history obtained in the hydrodynamical simulation is presented in 
Table~\ref{tab3}. The final kinetic energy is still on the low side but no 
longer very low whereas the amount of \element[][56]{Ni} (0.54~M$_{\sun}$) is enough to 
power the light curve. Nevertheless, there still remains 0.39~M$_{\sun}$ of 
unburnt carbon and oxygen, which should be compared with the 0.65~M$_{\sun}$ 
left at the end of the deflagration phase. The reason why there still remains 
this large mass of carbon and oxygen is due to both the shielding effect of the ashes produced during the deflagration phase and the degradation of the code resolution when the detonation propagated through low-density regions. 
However, even if a much better resolution were achieved we would not expect major changes in the total energy of the ejecta or in the mass of \element[][56]{Ni} synthesized.

\begin{table}
\caption[]{Nucleosynthesis products, in solar masses.}
\label{tab3}
\centering
\begin{tabular}{lcc|lcc}
\hline\hline
\noalign{\smallskip}
 & DDT3DA & DDT3DB & & DDT3DA & DDT3DB \\ 
\noalign{\smallskip}
\hline
\noalign{\smallskip}
\element[][12]{C}  & 0.14 & 0.19 & \element[][55]{Mn} & 1.2E-3 & 1.9E-3 \\
\element[][16]{O}  & 0.25 & 0.26 & \element[][54]{Fe} & 0.061 & 0.055 \\
\element[][20]{Ne} & 0.026 & 0.015 & \element[][56]{Fe} & 0.66 & 0.68 \\
\element[][22]{Ne} & 0.005 & 0.007 & \element[][57]{Fe} & 0.013 & 6.1E-3 \\
\element[][24]{Mg} & 0.014 & 0.008 & \element[][59]{Co} & 3.1E-4 & 1.4E-4 \\
\element[][28]{Si} & 0.064 & 0.047 & \element[][58]{Ni} & 0.059 & 0.055 \\
\element[][32]{S}  & 0.027 & 0.023 & \element[][60]{Ni} & 0.034 & 0.015 \\
\element[][36]{Ar} & 6.1E-3 & 5.6E-3 & \element[][61]{Ni} & 5.6E-4 & 2.2E-4 \\
\element[][40]{Ca} & 7.4E-3 & 6.7E-3 & \element[][62]{Ni} & 1.5E-3 & 5.6E-4 \\
\element[][44]{Ca} & 9.9E-5 & 2.3E-5 & \element[][65]{Cu} & 3.2E-6 & 1.4E-6 \\
\element[][48]{Ti} & 3.1E-4 & 1.9E-4 & \element[][64]{Zn} & 5.0E-4 & 1.9E-4 \\
\element[][52]{Cr} & 3.4E-3 & 2.7E-3 & \element[][66]{Zn} & 3.8E-5 & 1.4E-5 \\
\noalign{\smallskip}
\hline
\end{tabular}
\end{table}

\begin{figure}
\resizebox{\hsize}{!}{\includegraphics{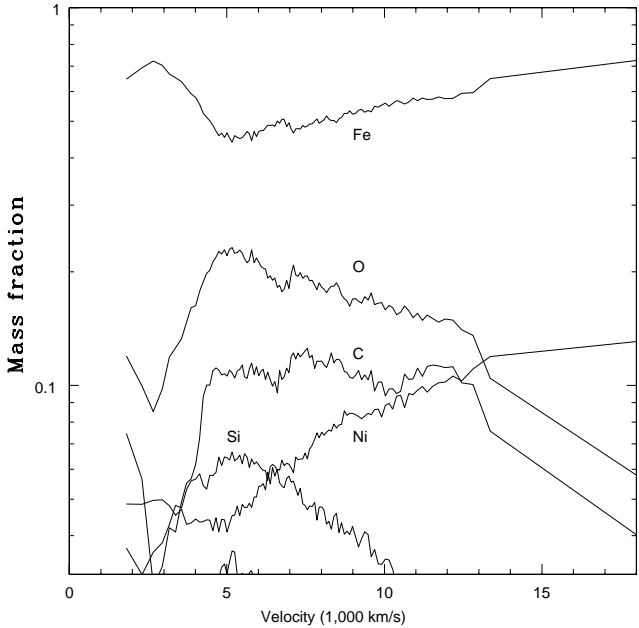}}
\caption{
Final distribution of elements in model DDT3DA as a function of velocity, after
radioactive decay.
}\label{fig10}
\end{figure}

The total amount of \element[][56]{Fe} (after radioactive decay) and intermediate-mass 
elements ejected in the explosion are 0.66~M$_{\sun}$~and 0.11~M$_{\sun}$ 
respectively, which is of the same order as the abundances deduced from the 
spectra of normal Type Ia supernovae and compatible with the demands of current models of galactic chemical 
evolution. Another positive feature of the model is that the ashes are not
concentrated in large clumps as is usually obtained in 
pure deflagration models. In addition, 
there remains little unburnt carbon and oxygen close to the center.
 Unfortunately, model DDT3DA does not completely remove all 
 nucleosynthetic deficiencies affecting current three-dimensional models of 
 SNIa. There are too many Fe-group elements close to the surface of the 
 white dwarf, especially too much radioactive \element[][56]{Ni} 
 (Fig.~\ref{fig9}). The distribution of species in velocity space 
(Fig.~\ref{fig10}) shows that the ejecta is not chemically
stratified, as demanded by recents observations of the light curve in the near 
and middle infrared \citep{hoe02} as well as by the 
properties of the X-ray emission of putative Type Ia supernova remnants 
\citep{bad06}. The Fe that makes up most of the ejecta (Fig.~\ref{fig10}, see
also Fig.~\ref{fig12}) is mostly produced during the explosion in the form of
radioactive \element[][56]{Ni}, whereas the curves labelled as 'Ni' trace
the distribution of stable iron-group species. Such stable
nuclei are nearly absent from the low velocity layers of the ejecta, which does
not match the requirements from infrared observations of SNIa 
\citep{mot06,ger07,fes07}.

\subsection{Sensitivity studies}

We have explored the sensitivity of our results to the parameters of the 
initial model and to the deflagration-to-detonation transition criterion. 

To start with, we changed the wavenumber of the perturbation imposed to the 
velocity field at the flame location at the beginning of the three-dimensional calculation. Decreasing the wavenumber from $l=20$ \footnote{Note that, at $t=0$, the wavelength of the perturbation was $\sim5$ times the resolution of the code for our reference model with $l=20$} to $l=7$ has virtually no
effect on the results of the simulation, as the global figures change by less than 2-3\%. Next, we checked the sensitivity to the amplitude of the 
perturbation. Although this parameter is more influential than the wavenumber, 
its imprint is still quite modest. Reducing the amplitude from 200 to 
100~km~s$^{-1}$ translates into a less energetic explosion (0.16 foes less than
in DDT3DA) and produces 0.11~M$_{\sun}$ less of \element[][56]{Ni}. Of course, this implies that the remaining mass of unburnt carbon and oxygen is larger than in DDT3DA. 

To explore the sensitivity of the results to the DDT criterion we took the same configuration at the end of the deflagration phase as in DDT3DA, but used alternative prescriptions to select the location of the induced detonation:
\begin{itemize}
\item Core detonation. All flame particles whose radius was lower than $1.3\times10^8$~cm were instantaneously incinerated. The density in these regions was higher than $\sim4\times10^7$~g~cm$^{-3}$.
\item Mid-altitude detonation. In this variant all flame particles within radii 
$1.7\times10^8$~cm and $2.0\times10^8$~cm were instantaneously incinerated. Their mean density was approximately $\sim2\times10^7$~g~cm$^{-3}$.
\item Atmospheric detonation. All flame particles within $2.4\times10^8$~cm and $2.8\times10^8$~cm were instantaneously incinerated. Their densities ranged from $\sim7\times10^6$~g~cm$^{-3}$ to $\sim10^7$~g~cm$^{-3}$.
\end{itemize}
The diversity of supernova explosions obtained with these prescriptions is not 
as rich as that usually achieved in one-dimensional delayed detonations. The 
kinetic energy varies at most in 0.15~foes, whereas the range of \element[][56]{Ni} masses 
spans from 0.34 to 0.43~M$_{\sun}$. By contrast, one-dimensional models with 
transition densities in the same range as above produce explosions with 
kinetic energies that change by 0.34~foes and \element[][56]{Ni} masses ranging from 0.38 to 0.97~M$_{\sun}$ \citep{bad03,bad05b}. Other 
properties of these three-dimensional explosions are similar to 
what was found in DDT3DA: the remaining mass of unburned carbon and oxygen is in all cases too large, and they lack the desired level of chemical stratification.

\subsection{Model DDT3DB}

The evolution of model DDT3DB during the first 0.81~s belonging to the deflagration phase
was identical to that of model B30U in \citet{gar05}. At that time a DDT was 
artificially induced in the vicinity of the flame. The properties of the model 
at DDT are given in Table~\ref{tab2}. In comparison with the configuration of 
DDT3DA at the moment of transition to detonation, the structure of 
DDT3DB was more centrally condensed, the central density being a factor $\sim4$ 
higher and the average flame density nearly $3.7\times10^7$~g~cm$^{-3}$. The
total energy and incinerated mass are, however, quite similar in both models, as
is the amount of unburned fuel. The major difference was in the mass of 
\element[][56]{Ni} synthesized during the deflagration phase, due to the longest time 
spent by DDT3DA at high densities, which enabled incinerated matter to
experience more electron captures. Consequently, the deflagration phase of model
DDT3DB was able to synthesize more \element[][56]{Ni}, but less neutronized
elements. This difference was still visible in the final results, after the
detonation phase, although somewhat attenuated (Table~\ref{tab1}). 

\begin{figure*}
\centering
\includegraphics[width=17cm]{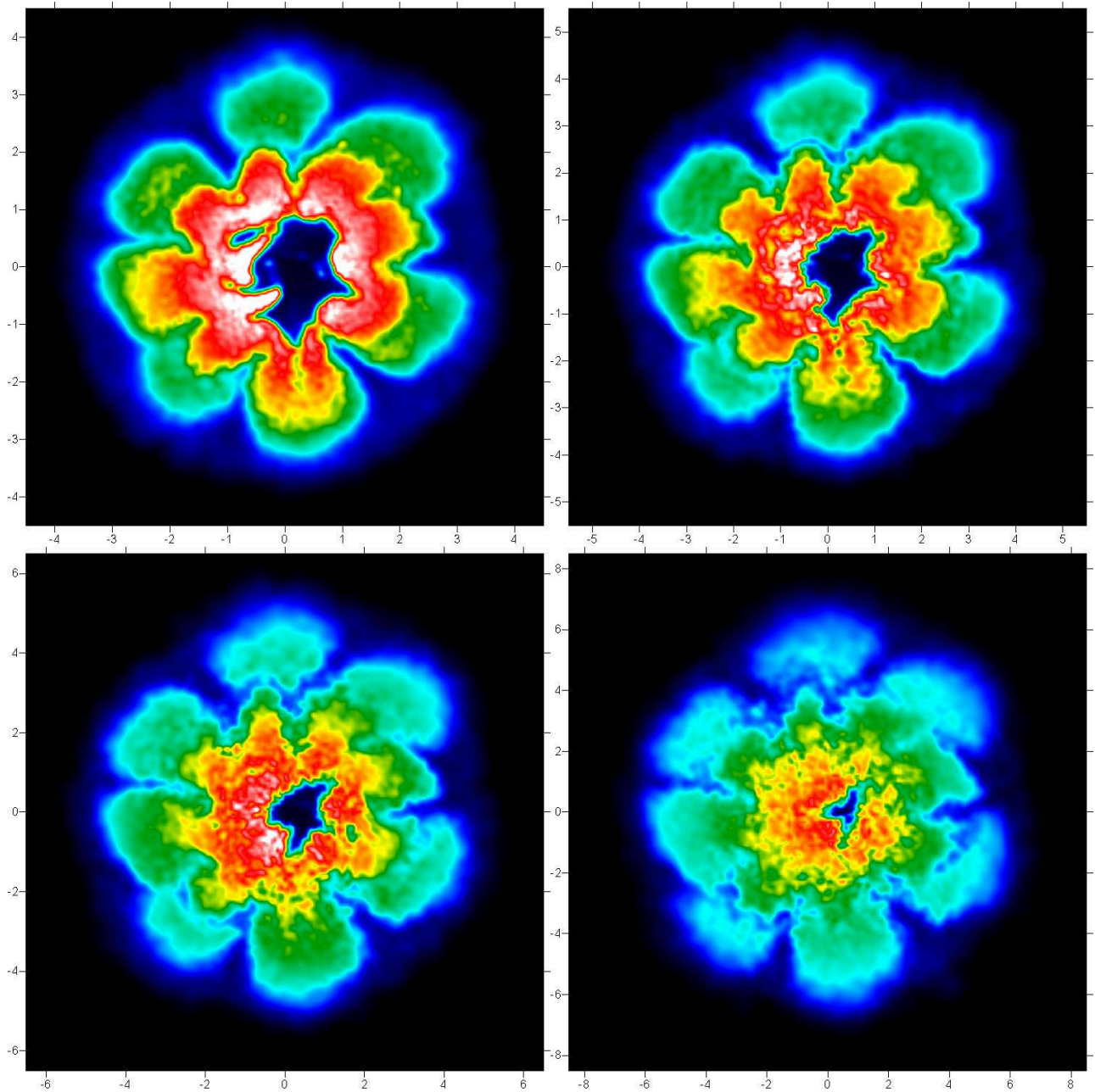}
\caption{
Snapshots of the temperature map in a cut across the star during the 
detonation phase of model DDT3DB, at times $t=0.8$, 0.9, 1.0, and 1.1~s. 
The scales of length and temperature are the same as in Fig.~\ref{fig1}.
}\label{fig12}
\end{figure*}

The temporal evolution of the detonation phase in model DDT3DB is shown in
Fig.~\ref{fig12}. As can be seen in Fig.~5 of \citet[][central figure in the
right column]{gar05} at the time of DDT the hot nuclear ashes formed a broken 
ring that enclosed a central volume rich in carbon and oxygen. The ensuing 
detonation preferentially propagated inwards 
processing most of the inner fuel into Fe-peak nuclei without 
having to deal with major obstacles composed of almost inert matter. However, 
a detonation wave was not successfully launched outwards and, as a result, the 
already vigorous expansion of the white dwarf prevented the outer layers from 
being processed by the nuclear bomb. 

\begin{figure}
\resizebox{\hsize}{!}{\includegraphics{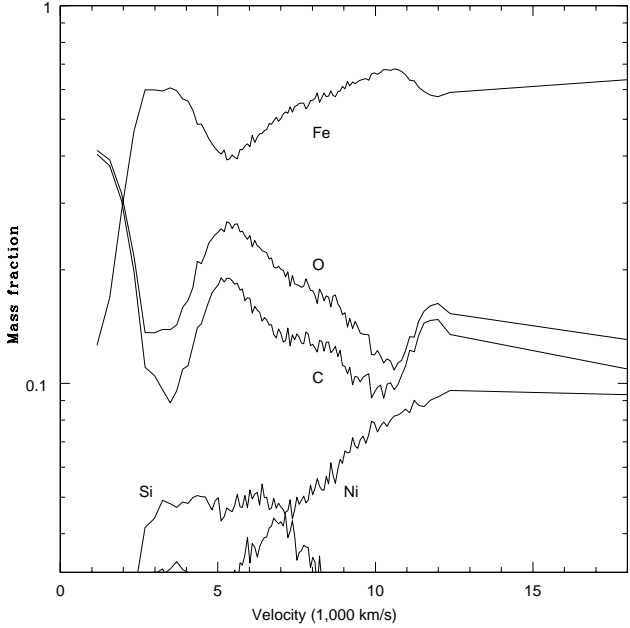}}
\caption{
Final distribution of elements in model DDT3DB as a function of velocity, after
radioactive decay.
}\label{fig14}
\end{figure}

Model DDT3DB shares the same problems as DDT3DA: low final kinetic 
energy, large mass of unburned carbon and oxygen, and lack of chemical 
stratification (Fig.~\ref{fig14}). A significative difference between both 
delayed detonation models is the amount of neutronized Fe-peak nuclei ejected 
in the explosion. The mass of slightly neutronized nuclei such as \element[][54]{Fe} 
and \element[][58]{Ni} was of the same order in both models because their production 
is determined by the initial presence of \element[][22]{Ne} (1\% in mass)
mixed with \element[][12]{C} and \element[][16]{O}, rather than by electron captures on NSE 
matter. The final ratio of the masses of \element[][54]{Fe} and \element[][58]{Ni} with respect 
to \element[][56]{Fe} is about 1.5 and 2 times solar respectively. However, the 
production of more neutronized Fe-peak nuclei in model DDT3DB is more scarce 
than in DDT3DA due to the rapid migration of the bubbles to 
low density regions shortly after the beginning of the deflagration phase. 
Had our initial models started from even higher central densities this effect 
would had been accentuated. 

\section{Conclusions}

The multidimensional delayed detonation model still remains an alternative way
to account for the observational properties of Type Ia supernovae, 
although there still remain many pieces of the puzzle to be settled. Among them 
is the mechanism of transition from 
deflagration to detonation, especially its viability in an exploding white 
dwarf. Although the most satisfactory approach to the problem would be to 
reveal the mechanism from first physical principles, this is an exceedingly 
difficult task in view of the many different length and temporal scales 
involved and of the huge diversity of initial conditions that can be imagined. 
Even though some advances have been brought about
\citep{kho91,kho97,nie97,nie99,lis00,rop04,zin07}
 no sound mechanism for the transition has yet been found. 

Given the lack of a consistent physical picture of the DDT mechanism 
the opposite path is usually taken, trying to constrain the conditions of DDT by 
comparing the resulting explosion with the observational data of SNIa. However, this approach is encumbered by the vast set of DDT conditions that need to be tested if a three-dimensional model has to be built. In this paper, we have tried to contribute to the still scarce set of three-dimensional 
simulations of delayed detonation models with two simulations in which the 
transition takes place at densities in the same range as those favored by 
succesful one-dimensional delayed detonation models of SNIa, i.e. $\rho_\mathrm{tr}\sim1-4\times10^7$~g cm$^{-3}$. 

Our delayed detonation models improve on the pure deflagration ones, but there 
still remain severe deficiencies compared to observations of SNIa, the two most important being the lack of chemical stratification and the huge mass of unburned carbon and oxygen ejected in the supernova explosion. While the amount of ejected carbon and oxygen might be affected by resolution issues, it 
is unclear if it can be lowered as much as needed to fit the observations. On 
the other hand, the lack of chemical stratification may not depend as much on
the details of the numerical implementation. To our knowledge, a chemically
stratified three-dimensional delayed detonation has only been found in a single work so far \citep{gam05}, but they had to assume a singularly high density of DDT.
The presence of high-velocity nickel above the silicon layer could also be at 
odds with the observations of SNIa, but this cannot be ascertained in the
absence of a three-dimensional spectral calculation.

Another interesting result is the relative insensitivity of the 
global properties of the explosion with respect to variations in the initial 
model (geometry of the runaway volume: centered vs bubbly) and location of the 
DDT (at a fixed transition time). Differences in kinetic energy of 0.15 foes 
and ejected masses of \element[][56]{Ni} of 0.09~M$_{\sun}$ as we obtained are in clear 
contrast to the rich diversity of explosions obtained with one-dimensional and two-dimensional (Arnett \& Livne 1994) delayed detonation models in which the DDT spanned similar density ranges. 

Model DDT3DB can be compared with the simulations performed by 
\citet{rop07}. In this last work three delayed detonation models were presented 
in which the initial runaway took place in different numbers 
of hot bubbles, whereas the densities of transition to detonation 
ranged from 1.3 to $2.4\times10^7$~g cm$^{-3}$, compared with
the transition density of model DDT3DB, $\rho_\mathrm{tr}=3.7\times10^7$~g
cm$^{-3}$ (Table~\ref{tab2}). In spite of having used
diverse numerical techniques and slightly different physics in both 
calculations, the final kinetic energy of our model DDT3DB (0.81 foes) fits 
quite well in the trend of the asymptotic kinetic energies given by R\"opke 
\& Niemeyer (1.5 to 1.0 foes). The same can be said about the ejected mass of 
unburned carbon and oxygen, which goes from 0.04 to 0.22~M$_{\sun}$ in their 
models, to be compared with 0.45~M$_{\sun}$ of C-O ejected in DDT3DB. The 
convergence of results of these different simulations supports 
the models reported here. 

\begin{acknowledgements}
We thank the referee for a careful reading of the manuscript and many positive
suggestions that helped to improved the clarity of the text. 
This work has been partially supported by the MEC grants AYA2005-08013-C03, 
AYA2004-06290-C02, by the European Union FEDER funds and by the Generalitat de 
Catalunya
\end{acknowledgements}

\bibliographystyle{aa}
\bibliography{ebg}

\end{document}